# INCA – Inductively coupled array discharge

**Philipp Ahr, Tsanko V. Tsankov, Jan Kuhfeld, and Uwe Czarnetzki**

Institute for Plasma and Atomic Physics, Ruhr University Bochum, 44780 Bochum, Germany

E-mail: `philipp.ahr@rub.de`

**Abstract.** Recently a novel concept for collisionless electron heating and plasma generation at low pressures was proposed theoretically (U Czarnetzki and Kh Tarnev, *Phys. Plasmas* **21** (2014) 123508). It is based on periodically structured vortex fields which produce certain electron resonances in velocity space. A more detailed investigation of the underlying theory is presented in a companion paper (U Czarnetzki, submitted to *Plasma Sources Sci. Technol.* (2018), arXiv:1806.00505).

Here, the new concept is experimentally realized for the first time by the INCA (INductively Coupled Array) discharge. The periodic vortex fields are produced by an array of small planar coils. It is shown that the array can be scaled up to arbitrary dimensions while keeping its electrical characteristics. Stable operation at pressures around and below 1 Pa is demonstrated. The power coupling efficiency is characterized and an increase in the efficiency is observed with decreasing pressure. The spatial homogeneity of the discharge is investigated and the behaviour of the plasma parameters with power and pressure are presented. Linear scaling of the plasma density with power and pressure, typical for conventional inductive discharges, is observed. Most notably, the plasma potential and the corresponding mean ion energy show clear evidence for the presence of super energetic electrons, attributed to stochastic heating. In the stochastic heating mode, the electron distribution function becomes approximately Maxwellian but with increasing pressure it turns to the characteristic local distribution function known from classical inductive discharges. Large area processing or thrusters are possible applications for this new plasma source.

## 1. Introduction

Inductively coupled plasmas (ICP) belong to the most versatile plasma sources at low pressures ($p \leq 100\,\mathrm{Pa}$) [1–3]. In these discharges energy is coupled directly to the electrons and therefore ICPs can perform with very high efficiency. In the simplest case, a flat coil serves as an antenna which induces an electric field in the plasma. The coil is usually driven with a sinusoidal current oscillating in radio frequency (RF) range, i.e. between $100\,\mathrm{MHz}$ and a few $10\,\mathrm{MHz}$, with the most common frequency being $13.56\,\mathrm{MHz}$ [2]. The induced electric field then drives an oscillating current in the plasma. The field and the associated current are limited to a region of typically a few cm in front of the antenna, called the skin depth. Within this region, the electric field amplitude decays exponentially. Energy is transferred from the electric field to the electrons by ohmic heating, i.e. via elastic collisions with the background neutral gas atoms or molecules [1]. At decreasing pressure, i.e. large mean free paths, ohmic heating becomes increasingly inefficient (typically less than $1\,\mathrm{Pa}$). In this case, still heating can be possible under certain conditions by a warm plasma effect called classical stochastic heating. There heating results from the inhomogeneity of the oscillating electric field on one hand and the thermal motion of the electrons in and out of the skin region on the other hand. This non-local collisionless or stochastic heating effect requires that the mean residence time of the electrons in the skin layer is short compared to the RF period [1, 4].

The above discharge concept is applied in research and industry for long. However, conceptionally, ICPs have a couple of shortcomings. Firstly, upscaling is difficult for several reasons. The flat coil antenna produces a doughnut shaped electric field and homogeneity can become a serious issue at large antenna sizes. Further, the antenna usually needs to be separated from the plasma by a quartz window in order to suppress capacitive coupling, i.e. the window serves as a capacitive voltage divider and reduces the potential over the sheath in front of the antenna. For large sizes the window needs to have a high thickness in order to withstand the pressure force from the atmosphere on the outer side. The corresponding large distance between the antenna coil and the plasma can reduce substantially the field strength available for electron heating. Typically this limits the antenna size to a diameter of less than about $30\,\mathrm{cm}$.

Several alternative antenna designs have been developed over the years. Solenoid or dome shaped antennas are probably the most common alternatives [2,3,5]. Further, a couple of arrays consisting of smaller coils have been proposed. There the idea is to distribute a number of conventional small antennas over a larger area. The latest development is probably the replacement of the coils by linear rods [6–8]. Apparently, a large number of parallel linear rods in a plane or along the walls of a cylinder are a simple way of upscaling. The drawback is a rather high current required to compensate for the relatively low field produced per rod. This is realized by a resonant network which produces standing waves which can have the required high current amplitudes.

In all cases electron heating is either ohmic or produced by an uncontrolled warm plasma effect. Recently an alternative concept for low-pressure collisionless heating was proposed by Czarnetzki and Tarnev [9]. The discharge concept provides a tailored heating mechanism with the inherent property of basically unlimited upscaling. It was shown theoretically that a planar array of periodic vortex field structures with fixed phases between the individual vortices provides certain resonances in velocity space for electron acceleration. It can be shown that the periodic vortex electric field of a phase correlated array can be decomposed into a number of plane waves pointing in different spatial directions with perpendicular electric fields. Naturally, the directions of the various wave vectors as well as the directions of the corresponding electric fields all lie in the plane of the array. The wavelength of these waves is identical to the size of an individual vortex and the frequency is determined by the applied RF frequency. Therefore, no dispersion relation exists and these waves might be better called pseudo waves. In any case, the waves have a well-defined phase velocity which is equal to the ration of the wavelength to the frequency. Electrons come into resonance when they move into the direction of the associated wave vector with a speed equal to the phase velocity. Since the field is oriented in a perpendicular direction, resonant acceleration and energy gain of an electron has no effect on the resonance condition, i.e. on the velocity along the wave vector. This ideal behaviour applies to a single resonance but in general electron trajectories resulting from multiple resonances, which are pointing in different directions, are more complex. Nevertheless, there is a substantial collisionless non-local heating across the plane of the array. Like in the classical stochastic heating case, electrons move in and out of the heating range within the skin depth by their thermal motion. The phase velocity of the array is a design parameter which is chosen to be a little smaller than the typical thermal velocity of the electrons. It should be noted that the resonances are not affected by the plasma density.

The idea is further developed in a recent theoretical work [10] which calculates the mean power per area transferred to the electrons as well as the complex susceptibility of the array and the related electric field propagation into the plasma. An important insight from this work is that the array structure provides non-local heating in the plane but is local vertical to the plane. Contrarily, classical stochastic heating is local in the plane but non-local vertically to the plane. The transition between these extremes depends on the ratio of the size of an individual vortices and the skin depth. With proper scaling, clearly the stochastic heating in the plane dominates.

In this work, the new concept of an inductively coupled array (INCA) is successfully demonstrated experimentally for the first time. The proposed periodic vortex field structure is realized by an antenna consisting of an array of 36 individual small coils. A discharge utilizing this antenna array is then investigated and its performance is characterized through various diagnostics. Where possible, comparison with the theory is made. Measurements of the RF current through the coils as a function of the generator power together with a simple model allow the power coupling efficiency to be estimated. It reaches 50 % in the lower pressure range investigated, with further optimization being possible. The plasma transits to the H-mode already at relatively low power levels (100 W at 1 Pa). The plasma density plasma shows the usual diffusion-controlled spatial profile. The electron energy distribution function (EEDF) is non-local and Maxwellian. Evidence is found for the presence of super energetic electrons (energies exceeding 40 eV). The plasma density in the discharge centre exhibits a linear dependence on the power and the pressure in agreement with theoretical expectations and results in conventional inductive discharges. Overall, the experimental results support the theoretically predicted feasibility and good efficiency for plasma production at low pressures (at and below 1 Pa) of a discharge based on the concept of Czarnetzki and Tarnev [9].

It has to be also noted, this is only the initial investigation of the source. Several aspects can be optimized even further and a number of open questions still remain. Most of them can profit from a dedicated modelling efforts. Another aspect that is stressed in this work is the possibility for easy scalability of this discharge. Through the analysis of the antenna it is shown that it is scalable to large dimensions while keeping the electric characteristics of the matching circuit unchanged.

The characteristics of the INCA discharge are rather suitable for a number of applications. Scalability makes the concept attractive for large area plasma processing, as the one required in e.g. flat panel production or surface treatment. The efficient operation at low pressures is well suited for designs of novel plasma-based space propulsion. The high values of the plasma potential due to the super energetic electrons are also beneficial for these applications.

The paper is structured as follows. First the experimental setup including some design considerations of the antenna array are presented together with the diagnostic methods used. Next, the electrical characteristics and the scalability of the antenna array are discussed. Then the experimental results are presented and analysed. Here, the transition to the H-mode, the spatial profiles of the plasma parameters and their variation with the power and the pressure are discussed. Finally, the work is summarized and some conclusions are drawn.

## 2. Experimental setup

In this section, the antenna array, the general setup and the various diagnostics are described. The design considerations and the construction of the antenna array are emphasized, since the antenna array plays a central role in the production of an electric field with the required specially tailored spatial structure.

### 2.1. Antenna array

The antenna array is sketched in figure 1(a) and a photograph is given in figure 3(b). It consists of an array of $6 \times 6$ small identical planar coils. Each coil is a two-turn spiral with an outer diameter of 46 mm. Considering also the spacing between the individual coils, the spatial constant of the array, i.e. its cell size [10], is $\Lambda = 50$ mm.

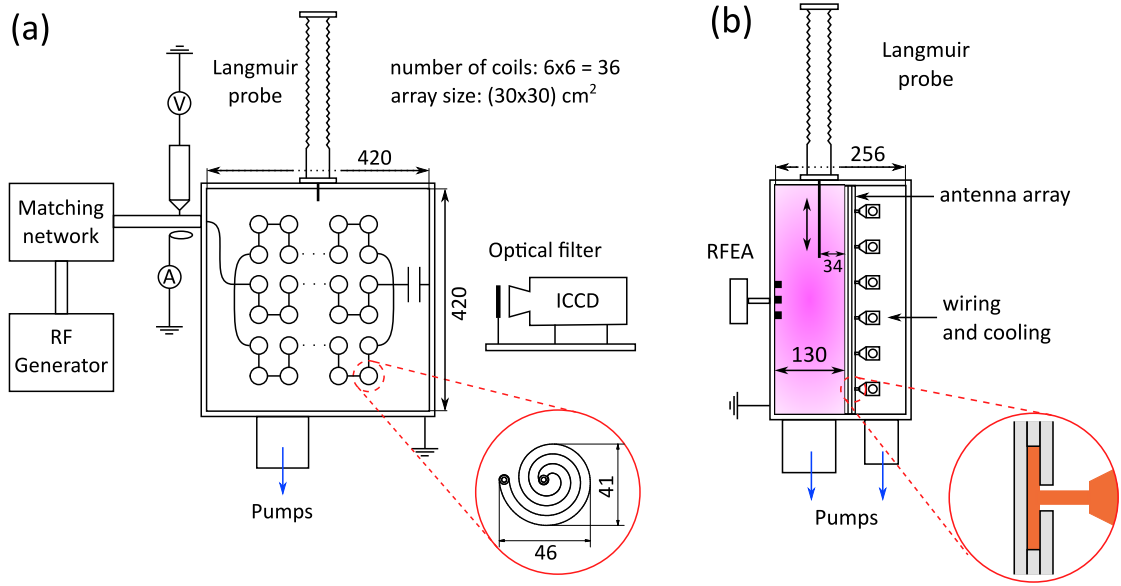

**Figure 1.** Schematics of the setup in a front view (a) and a side view (b). The dimensions are given in mm.

The corresponding phase velocity of the array is $v_{\mathrm{ph}} = \Lambda\,\omega_{\mathrm{rf}} = 0.68 \cdot 10^6\,\mathrm{m/s}$ which is smaller than the typical electron velocity of $10^6\,\mathrm{m/s}$. Here $\omega_{\mathrm{rf}}$ is the driving frequency of the array (for $f_{\mathrm{rf}} = 13.56\,\mathrm{MHz}$, $\omega_{\mathrm{rf}} = 8.52 \cdot 10^7\,\mathrm{s}^{-1}$).

The spacing between the turns and the width of the copper windings are both $5\,\mathrm{mm}$. The $3\,\mathrm{mm}$-thick coils are placed in a glass mask of the same thickness, that holds them in place. The coils and the mask are placed between a glass plate on one side and a quartz window on the side facing the plasma, both with a thickness of $4\,\mathrm{mm}$. This sandwich structure (figure 1(b)) then holds the individual parts of the antenna array together. On the back side, the coils are connected to thick copper clams, fixed on water cooled ceramic tubes (figure 3(a)). Therefore, cooling of the coils is provided by heat conduction. This turned out to be an efficient cooling mechanism even at powers of $1\,\mathrm{kW}$. An important advantage of this passive cooling system is the independence of cooling and electrical connection, which is an unusual approach. Here, the coils are connected with thin copper stripes (see also figure 3(a)), which enables effortless variation of the electrical wiring.

In the experiment, the coils in the whole array are divided into three parallel arms in each of which twelve coils are connected in series. In the general case, the antenna array can be divided into $p$ arms with $q$ coils in each arm, e.g. in the experiment $p = 3$ and $q = 12$. The antenna array then consists of $p \times q$ coils, each one having the complex impedance $\hat{Z}_{\mathrm{c}}$. The impedance of one of these arms is $\hat{Z}_{\mathrm{arm}} = q\hat{Z}_{\mathrm{c}}$ and the overall impedance of the array is $\hat{Z}_{\mathrm{a}} = \hat{Z}_{\mathrm{arm}}/p = (q/p)\hat{Z}_{\mathrm{c}}$. Apparently, changing the array in such a way that $q/p$ remains constant would not change the impedance of the antenna array and, therefore, also the matching conditions.

Each coil in the antenna array has an inductance of $L_{\mathrm{c}} = (101 \pm 10)\,\mathrm{nH}$. This value has been obtained by incorporating one coil into a resonance LC-circuit (with a capacitance of $C = 10\,\mathrm{nF}$) and measuring the resonance curve yielding the resonance frequency ($f_{\mathrm{res}} = 5\,\mathrm{MHz}$). The value for the inductance is in good agreement with theoretical estimations based on the geometrical dimensions of the coil [11,12]. Without plasma, the impedance of an individual coil is mostly reactive $\hat{Z}_{\mathrm{c}} \approx i\omega_{\mathrm{rf}}L_{\mathrm{c}}$. Then the calculated overall complex impedance of the antenna array is $\left|\hat{Z}_{\mathrm{a}}\right| = 4\omega_{\mathrm{rf}}L_{\mathrm{c}} = (34 \pm 3)\,\Omega$. Naturally, this value is slightly underestimated, since it does not account for the mutual inductance between the coils and the contributions of the wiring. The entire antenna array is connected to a vacuum feedthrough. On the atmospheric pressure side of the feedthrough, a voltage and current measurement systems are installed. Measurements on this entire setup, performed with a network analyzer (Agilent 8714 ES RF 300kHz 3GHz), give $\omega_{\mathrm{RF}}L_{\mathrm{a}} = 44\,\Omega$ (corresponding to an actual antenna inductance of $L_{\mathrm{a}} = 520\,\mathrm{nH}$ instead of the calculated $(404 \pm 40)\,\mathrm{nH}$) with an additional ohmic resistance of $R_{\mathrm{a}} = 0.6\,\Omega$. This ohmic resistance is attributed mostly to the vacuum feedthrough and the current and voltage measurement probes since this portion of the setup heats up during high power operation. However, in the following no distinction will be made between the electrical characteristics of the antenna array itself and of the vacuum feedthrough. Instead, they are grouped together for simplicity as a single unit under the term antenna array.

To decrease the capacitive coupling between the coils and the plasma, i.e. the voltage drop over the coils, a vacuum capacitor (Comet) with $C_{\mathrm{g}} = 500\,\mathrm{pF}$ is connected in series between the array and the ground connection (figure 1(a)). This value is chosen such that $\omega_{\mathrm{rf}}L_{\mathrm{a}} \approx (\omega_{\mathrm{rf}}C_{\mathrm{g}})^{-1}$. This provides optimal conditions for the suppression of the capacitive component of the power coupling [13], which permits the transition to the H-mode to be already at about $100\,\mathrm{W}$ (section 3.3). The final impedance of the antenna array is measured to be $\hat{Z}_{\mathrm{t}} = (0.6 + 22\mathrm{i})\,\Omega$. As mentioned, the real part of this impedance is due to the vacuum feedthrough and current and voltage probes. Indeed, this causes significant power losses and limits the power coupling efficiency (see section 3.1). On the positive side, the changes in the antenna resistance due to the presence of the plasma are reduced which allows optimal matching within wider range of the discharge parameters (power and pressure). However, it is apparent that future improvements and optimizations are possible.

For a matching point to exist, certain criteria on the resistance and the inductance of the array have to be fulfilled. The matching conditions for an L-type matching network (figure 2) with a load impedance $\hat{Z}_{\mathrm{t}} = R_{\mathrm{t}} + iX_{\mathrm{t}}$ are [1]:

$$\omega_{\mathrm{rf}}C_1 R_0 = \left(\frac{X_{\mathrm{t}}}{R_0} - \sqrt{\frac{R_{\mathrm{t}}}{R_0}}\sqrt{1 - \frac{R_{\mathrm{t}}}{R_0}}\right)^{-1} \approx \frac{R_0}{X_{\mathrm{t}}} = \frac{R_0}{\omega_{\mathrm{rf}}L_{\mathrm{a}} - \frac{1}{\omega_{\mathrm{rf}}C_{\mathrm{g}}}}, \qquad (1)$$

$$\omega_{\mathrm{rf}}C_2 R_0 = \sqrt{\frac{R_0}{R_{\mathrm{t}}} - 1}. \qquad (2)$$

Here $R_0 = 50\,\Omega$ is the output impedance of the generator and $\hat{Z}_{\mathrm{t}}$ is the transformed impedance of the antenna, i.e. its impedance when the plasma is present.

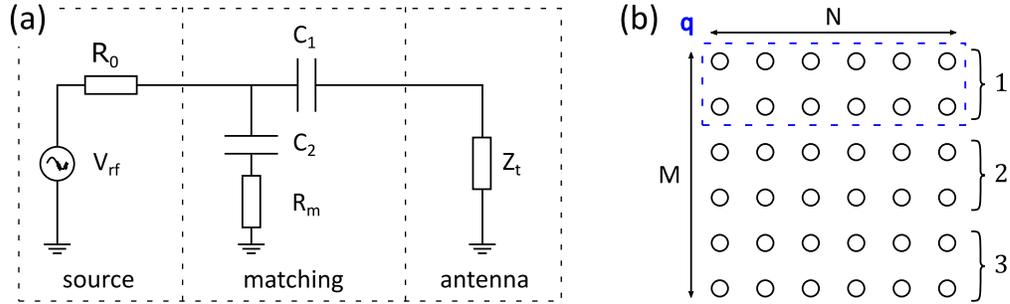

**Figure 2.** (a) Schematic representation of an L-type matching network. (b) Grouping of the individual coils for optimal wiring.

The requirement $C_2 \geq 0$ sets the constraints on the impedance $\hat{Z}_t$ that can be matched: $0 \leq R_t \leq R_0$. In the present setup $R_t \ll R_0$ for all investigated plasma conditions. This allows the approximation in (1). It shows that the position of the adjustable capacitor $C_1$ is essentially independent from the transformed ohmic resistance $R_t$ of the antenna array. Consequently, during plasma operation the capacitor $C_1$ is varied only in a narrow range, since $X_t$ is only weakly effected by the plasma [14]. In contrast $C_2$ varies strongly with the plasma conditions.

## 2.2. Discharge Chamber

The plasma is ignited in a stainless steel vacuum chamber with inner dimensions of $420\,\mathrm{mm} \times 420\,\mathrm{mm} \times 256\,\mathrm{mm}$ (figure 1). The antenna array is positioned nearly in the middle of this volume and divides the chamber into two separate volumes, sealed from each other. On one side the plasma is ignited and the other side houses the antenna wiring and water cooling. Due to the large surface area and the small thickness of the plasma-facing quartz window of $4\,\mathrm{mm}$, the antenna sandwich structure cannot withstand large pressure forces. Evacuating the back side of the array avoids this problem. The two volumes are evacuated by separate turbomolecular vacuum pumps combined with dual diaphragm fore-pumps. A pressure equilibrating system is installed to ensure that the pressure difference on both sides does not exceed $1000\,\mathrm{Pa}$.

The plasma side has a depth of $130\,\mathrm{mm}$, resulting in a plasma volume of about $23\,\ell$. This volume is pumped by a turbomolecular pump (TMU 521 P, Pfeiffer Vacuum), providing a base pressure of about $10^{-5}\,\mathrm{Pa}$. An adaptive pressure-controlled valve ensures a constant working pressure during operation. The working gas is filled in the volume by four gas inlets, embedded in the sidewalls. The gas flow rate is controlled by a MKS mass flow controller. Operation pressure in the range of $0.1\,\mathrm{Pa}$ to $20\,\mathrm{Pa}$ can be chosen. In this study pure argon and neon (5.0) were used as working gases.

The second volume of the discharge chamber, the vacuum side, has a depth of $115\,\mathrm{mm}$, corresponding to a volume of about $18\,\ell$. The electrical wiring and the water cooling for the antenna are situated here (see figure 3(a)). This volume is pumped by another turbomolecular pump (EXT 70H, BOC Edwards), providing a base pressure better than $10^{-2}$ Pa (also during plasma operation). This pressure is sufficient to reduce the stress on the sandwich structure of the antenna and to prevent discharge ignition during plasma operation.

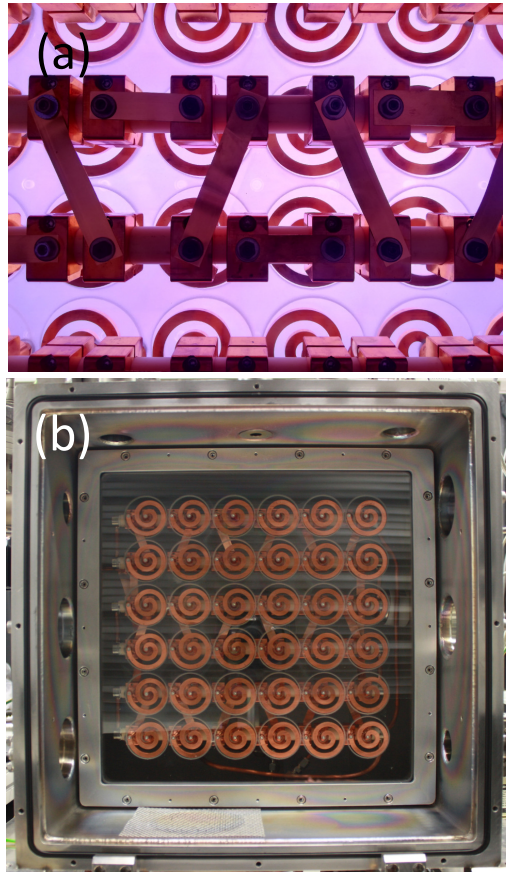

**Figure 3.** Photographs of the setup. (a) Back view during discharge operation in argon. Part of the coil wiring is also visible. (b) Front view showing the arrangement of the coils in the antenna array.

RF power at a frequency of 13.56 MHz is provided via a power supply (Dressler Cesar 1312) and an L-type matching unit. The matching unit consists of two tunable vacuum capacitors (Comet (5-500) pF), one connected to ground, parallel to the antenna array and the other one in series with the antenna. A fixed capacitor with 500 pF is connected in parallel to the tunable capacitance connected to ground. The matchbox is directly connected to the current-voltage measurement unit which itself is connected to the vacuum feedthrough.

### 2.3. Diagnostics

Several diagnostics are used to monitor and characterize the plasma performance. Firstly, a high-voltage probe (Tektronix P6015A, bandwidth 100 MHz) and a home-made inductive pick-up loop are installed near the vacuum feed-through to monitor the RF voltage and current to the antenna array. The pick-up loop is calibrated to obtain absolute values of the current using a voltage-current meter (Octiv, Impedans Ltd). The uncertainty in the values for the current are mostly due to this calibration. A calibration for the phase shift between the current and the voltage has been attempted. However the uncertainty ($\pm 4°$) was comparable to the deviation from 90° of the phase shift between current and voltage. Therefore, the power delivered to the antenna array was not directly measured, but is obtained through a simple model.

The ion energy at the walls is measured using a 3-grid retarding field energy analyzer (RFEA Semion, Impedans Ltd). The first derivative of the current-voltage characteristic is proportional to the ion velocity distribution function (IVDF) as a function of the energy. The obtained IVDFs exhibit a single peak because under the investigated conditions the wall sheath is collisionless. Therefore, the energy at which this peak occurs equals approximately the voltage drop over the sheath.

A home-made Langmuir probe is installed on the top side of the chamber (figure 1). The probe is positioned in the middle of this 420 mm-long side. The probe is movable in a plane parallel to the antenna array at a distance of 34 mm from the quartz window. A tungsten probe tip with a length of 6 mm and a radius of 0.025 mm is used. The size of the port for the Langmuir probe is reduced by an orifice with 13 mm diameter to minimize disturbances to the probe measurements in the vicinity of the chamber wall. Measurements were carried out at distances from the side wall between 1 cm and 22 cm (center point at 21 cm). Before each measurement the probe tip is cleaned by biasing it for 2 s at $-60$ V, i.e., 80-100 V below the plasma potential. The current-voltage (U-I) characteristics of the probe are taken in steps of 0.2 V. The recorded U-I characteristics are then analyzed by the Druyvesteyn method to obtain the electron energy probability function (EEPF). The second derivative of the probe current is obtained by a three-point numerical differentiation scheme with an adaptive step size. This method is used since it provides simultaneously also noise reduction and removes the need of further smoothing of the data.

The plasma potential is obtained from the zero crossing of the second derivative. The uncertainty in its values is 0.4 V and results from the voltage steps used to record the probe U-I characteristics. The electron density $n_e$ and the (effective) electron temperature $k_B T_e$ at a given position $x$ are calculated as integrals over the EEPF $f_p(x, \varepsilon)$ at that position:

$$n_e(x) = \int_0^\infty \varepsilon^{1/2} f_p(x, \varepsilon) \, d\varepsilon, \tag{3}$$

$$k_B T_e(x) = \frac{2}{3} \langle \varepsilon \rangle = \frac{1}{n_e} \int_0^\infty \varepsilon^{3/2} f_p(x, \varepsilon) \, d\varepsilon. \tag{4}$$

The uncertainty for these quantities is estimated as in [15]. This results in a typical uncertainty of the density of about 5 % and of the electron temperature of about 8 %.

Several view ports in the chamber walls allow for optical access. Three such view ports are available along the diagonal of the $420\,\text{mm} \times 420\,\text{mm}$ side opposite of the antenna array. An optical fiber with a collimating lens is used to collect the light at positions $1\,\text{cm}$ apart (within the view ports). The spectrum is then detected by a compact spectrometer (HR4000, OceanOptics). The spectral sensitivity has been calibrated by a calibrated Ulbricht's sphere.

Tunable diode laser absorption spectroscopy (TDLAS) is used to measure the metastable density and the neutral gas temperature. The absorption profile around the central wavelength ($763.51\,\text{nm}$) of the $2\text{p}_6 \rightarrow 1\text{s}_5$ transition (Paschen notation) of the argon atom is recorded. From the laser intensity with the RF power switched on ($I_\text{a}(\nu)$), and off ($I_0(\nu)$) the density of the lower metastable state of argon $1\text{s}_5$ is obtained:

$$n_m = \frac{4\varepsilon_0 m_\text{e} c}{f_{ik} L e^2} \int\limits_0^\infty \ln\left(\frac{I_0(\nu)}{I_\text{a}(\nu)}\right) \text{d}\nu. \tag{5}$$

Here $\varepsilon_0$ is the dielectric constant, $m_\text{e}$ the mass of an electron, $c$ the speed of light, $f_{ik}$ the oscillator strength of the absorption line ($f_{ik} = 0.214$ for the considered transition [16]), $L$ the length of the absorbing medium, $e$ the elementary charge and $\nu$ the frequency of the light. Due to saturation of the absorption profile, measurements are possible only within a limited range of conditions. Within this range the values of the metastable density are between $3 \cdot 10^{16}$ and $6 \cdot 10^{16}\,\text{m}^{-3}$ with an uncertainty of $5\,\%$. This uncertainty is mostly due to the uncertainty in the spatial profile of the metastable atoms along the absorption path.

At low pressures, the line broadening is dominated by the Doppler effect due to the thermal motion of the atoms. The width (FWHM) of the resulting Gaussian absorption profile, $\nu_\text{D}$, is then connected with the temperature of the neutral gas by

$$T_\text{g} = \frac{m_\text{a} c^2}{8 \ln(2) k_\text{B}} \left(\frac{\nu_\text{D}}{\nu_0}\right)^2, \tag{6}$$

where $m_\text{a}$ is the mass of the argon atom, $k_B$ is the Boltzmann constant and $\nu_0$ is the central frequency of the transition. Under all investigated conditions the gas temperature remains close to room temperature ($T_\text{g} = 300 - 350\,\text{K}$). The precision is about $2\,\%$ or $\pm 7\,\text{K}$ and is determined by the accuracy of the FWHM $\nu_\text{D}$.

Another view port (in the middle of the $420\,\text{mm}$-long side, figure 1(a)) is used to measure the modulation within the RF cycle of the emission intensity from the plasma. An intensified CCD camera (PicoStar HR16, LaVision) is positioned in front of the view port. A Nikon lens (AF NIKKOR, $f = 50\,\text{mm}$, focused at infinity) provides two dimensional images of the plasma emission in front of the antenna array. For the weak modulation of the intensity to be measurable, an emission line from a state with a lifetime much shorter than the RF period, $T_\text{rf}$, is needed (for $f = 13.56\,\text{MHz}$, $T_\text{rf} \approx 74\,\text{ns}$). Commonly the neon $2\text{p}_1 \rightarrow 1\text{s}_2$ line at $585.5\,\text{nm}$ is used [17]. This line has a lifetime of $14.5\,\text{ns}$ and is spectrally well isolated from other strong lines. Further, this transition fulfils also the second requirement – to be excited from the ground state, i.e. being sensitive only to the high energy tail of the electron distribution. An optical bandpass filter (central wavelength 585 nm, FWHM of 10 nm) is positioned in front of the objective of the ICCD camera to allow only the intensity of this neon line to be registered.

The emission intensity, $I(t)$, is recorded within 2 ns-wide windows in steps of 2 ns synchronous with the signal from the RF power generator. Then the ratio, $\chi(t)$, of the time-dependent intensity and the average intensity within one RF cycle, $\langle I(t) \rangle$, is formed:

$$\chi(t) = \frac{I(t)}{\langle I(t) \rangle}. \tag{7}$$

This ratio has the advantage of being insensitive to the spatial distribution of the electrons and the neutrals and is independent from the camera settings (exposure time, gain), optical transmittance (filter, objective, window) and camera sensitivity. However, the modulation of the intensity $\chi(t)$ still contains information on the oscillation of the electron energy which can be related to the strength of the induced electric field [18]. In inductive discharges, the mean energy of the electrons depends on the square of this induced field. Therefore, the energy of the electrons and, consequently, the emission intensity oscillate at twice the RF frequency $\omega_{\mathrm{rf}}$. However, when also capacitive coupling is present, the phase resolved modulation shows deviations from a pure harmonic oscillation at $2\,\omega_{\mathrm{rf}}$ and particularly exhibits a strong first harmonic.

To separate the two effects a Fourier analysis of $\chi(t)$ is performed. The complex amplitude $\hat{\chi}_{2\omega}$ at $2\,\omega_{\mathrm{rf}}$ is proportional to the square of the amplitude of the induced electric field. Then from the 2D information for the modulation of the emission intensity the spatial distribution of the induced electric field can be obtained:

$$E_{\mathrm{ind}} \propto \sqrt{|\hat{\chi}_{2\omega}|}. \tag{8}$$

The phase factor in $\hat{\chi}_{2\omega}$ contains information on the relative phase of the field:

$$\Phi_{\mathrm{ind}} = \frac{1}{2} \arctan\left( \frac{\Im(\hat{\chi}_{2\omega})}{\Re(\hat{\chi}_{2\omega})} \right). \tag{9}$$

Here $\Re$ and $\Im$ represent the real and the imaginary part of a complex number, respectively.

## 3. Results and discussions

In this section, the results from the various diagnostics are presented. The discussions are concentrated on the standard case of a discharge in argon at various pressures and power levels in the range of 0.5 to 20 Pa and 20 to 1000 W. However, where necessary, e.g., by the specifics of the diagnostics technique, results from discharges in pure neon are also presented. The discussions first concentrate on the electrical characteristics of the discharge, on its operation in the H-mode and on the distribution of the induced electric field in the plasma. Subsequently, the spatial characteristics of the plasma parameters and the distribution functions of the electrons are presented and their variation with power and pressure are outlined.

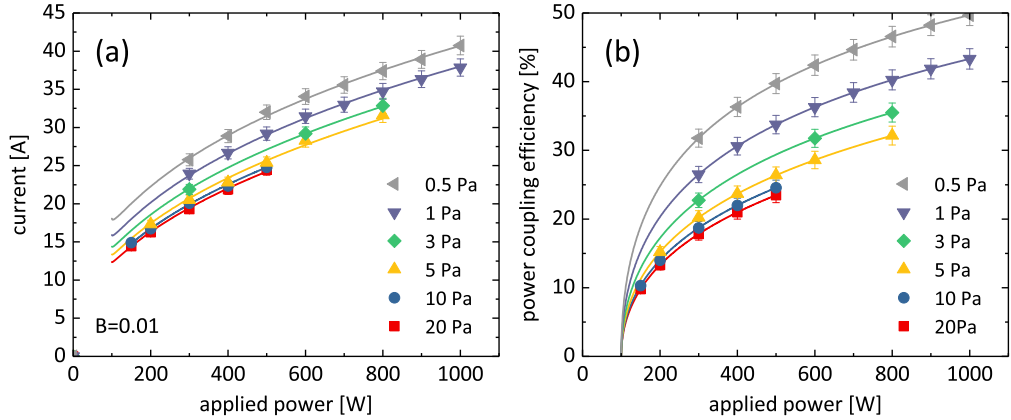

**Figure 4.** (a) Dependence of the current through the antenna array on the power from the RF generator for the case of a discharge in argon at various pressures. The solid curves are fits of equation (12) to the measured data points. (b) Power transfer efficiency $\eta$ calculated from (13) using the values of the fit parameters from (a). The errorbars are obtained from the uncertainties of the fits.

## 3.1. Electrical characteristics

The first aspect of the new plasma source that is investigated is its electrical characteristics. The measured current amplitude $I$ through the antenna array for a discharge in argon is shown in figure 4(a). The behaviour is similar also in discharges with neon. The current is of the order of few tens of amperes, which is a typical range for inductive discharges [19].

The current as a function of the power from the RF generator shows a nearly $P^{1/k}$ dependence (figure 4(a)). Measurements in a classical ICP discharge [19] reveal a value of $k = 1$. A model with a constant (i.e. power-independent) load would predict $k = 2$. However, the measurements here show that the value is closer to $k = 4$. To describe the observed trends the following considerations can be applied. The power deposited into the plasma, $P_{\mathrm{p}}$, is proportional to the product of the plasma density in front of the antenna $n_{\mathrm{s}}$, the skin depth $s$ and the square of the induced electric field amplitude $E$:

$$P_{\mathrm{p}} \propto s n_{\mathrm{s}} E^2. \tag{10}$$

This relation is in agreement with the classical expression in the case of ohmic heating [1]. Similar scaling is obtained also in [10] for the stochastic heating in the INCA discharge where it is also shown that the skin depth scales in the classical way like $s \propto n^{-1/2}$. As will be shown below (see section 3.5) the plasma density in the centre scales linearly with the applied RF power $n \propto (P_{\mathrm{rf}} - P_0)$, with $P_0 \approx 100\,\mathrm{W}$. By combining these relations and considering that the induced electric field strength is proportional to the amplitude of the current through the antenna, $E \propto I$, an expression for $P_{\mathrm{p}}$ is obtained: $P_{\mathrm{p}} = B\sqrt{P_{\mathrm{rf}} - P_0}I^2$, where $B$ is a proportionality constant.

Naturally, the power absorbed by the plasma, $P_{\mathrm{p}}$, is not identical to the power delivered by the RF generator, $P_{\mathrm{rf}}$. The difference is due to ohmic losses in the wiring, $P_{\mathrm{a}}$, and in the matching unit, $P_{\mathrm{m}}$. Generally, these losses are hard to quantify exactly. Here only approximates will be given.

The ohmic losses in the antenna can be estimated through the measured resistance of the antenna and its wiring $R_{\mathrm{t}} = 0.6\,\Omega$ as $P_{\mathrm{a}} = \frac{1}{2}R_{\mathrm{t}}I^2$. To quantify the losses in the matching unit, it will be treated as a current transformer with a small but finite resistance $R_{\mathrm{m}}$. The matching unit converts the small current from the RF generator ($I_{\mathrm{rf}} = \sqrt{2P_{\mathrm{rf}}/R_0} \leq 6\,\mathrm{A}$) into two large currents through the two arms of the matching, i.e. through $C_2$ to ground and through $C_1$ to the antenna (figure 2(b)). The latter current is the one being measured and presented in figure 4(a). Since these two currents are much larger than the current from the RF generator ($I \gg I_{\mathrm{rf}}$), their amplitudes can be approximated as being nearly equal. Then the losses in the matching unit are given by $P_{\mathrm{m}} = \frac{1}{2}R_{\mathrm{m}}I^2$.

The losses, described by $R_{\mathrm{m}}$, are mostly due to (mechanical) imperfections in the tunable capacitors. Therefore, the value of $R_{\mathrm{m}}$ is expected to vary with the settings of the capacitors. Experimentally, these settings are much less affected by the changes in the input power than by the variation in the pressure. Therefore, the same value of $R_{\mathrm{m}}$ should be applicable to all sets of measurements at the same pressure.

Certainly, also other models for the description of the power losses due to the matching network are possible. An example would be to consider these losses to be proportional to the RF power. However, the final functional form for the dependence between the current and the power from the generator (equation (11)) remains unchanged. Only the interpretation of the various coefficients is altered.

The power balance equation ($P_{\mathrm{rf}} = P_{\mathrm{m}} + P_{\mathrm{a}} + P_{\mathrm{p}}$) provides a relation between the measured current and the power delivered by the RF generator:

$$I = \sqrt{\frac{P_{\mathrm{rf}}}{(R_{\mathrm{t}} + R_{\mathrm{m}})/2 + B\sqrt{P_{\mathrm{rf}} - P_0}}}. \tag{11}$$

Apparently, in the two asymptotic cases the current reproduces the $P_{\mathrm{rf}}^{1/k}$-dependence: for very large values of $P_{\mathrm{rf}}$ ($P_{\mathrm{rf}} \gg P_0$, $[(R_{\mathrm{t}} + R_{\mathrm{m}})/(2B)]^2$) with $k = 4$ and for $P_{\mathrm{rf}} \to P_0$ with $k = 2$.

Expression (11) contains two known parameters, $P_0 = 100\,\mathrm{W}$ and $R_{\mathrm{eff}} = 0.6\,\Omega$, which are available from the experiment. The values of the other two parameters, $R_{\mathrm{m}}$ and $B$, can be obtained by fitting equation (11) to the experimental data. A more convenient form for the fitting is:

$$I = A\sqrt{\frac{P_{\mathrm{rf}}}{1 + A^2 B\sqrt{P_{\mathrm{rf}} - P_0}}}\,, \text{ with } A = \sqrt{\frac{2}{R_{\mathrm{t}} + R_{\mathrm{m}}}}. \tag{12}$$

**Table 1.** Values of the fitting parameters $A$ and $B$ in (12) and the corresponding values of $R_m$ in (11).

| $p$ [Pa] | 0.5 | 1 | 3 | 5 | 10 | 20 |
|---|---|---|---|---|---|---|
| $A$ [$1/\sqrt{\Omega}$] | 1.815 | 1.596 | 1.442 | 1.338 | 1.276 | 1.238 |
| $R_m$ [$\Omega$] | 0.01 | 0.18 | 0.36 | 0.52 | 0.63 | 0.70 |
| $B$ [$\sqrt{\Omega}/\text{A}$] | | | 0.01 | | | |

This expression has been fitted to the experimental data in figure 4(a). The fitted curves are also shown in the figure, but only for $P_{rf} \geq P_0 = 100\,\text{W}$ since for lower power values the discharge operates in the E-mode and equation (12) is not valid. The values of the fitting parameters – $A$ and $B$ – are given in table 1 together with the resulting values for the resistance $R_m$. Very good agreement with the experimental data and reasonable values for the various parameters are obtained. The parameter $B$ is independent of pressure and power (for a discussion on its independence on the pressure, see section 3.5). It has the value of $0.01\,\sqrt{\Omega}/\text{A}$. This parameter summarizes the various proportionality factors in (10), which allows also its theoretical estimation. Notably, the order of magnitude estimation is identical with the value from the fits.

The obtained good fits with a single value of the resistance $R_m$ is also in agreement with the weak changes in the matching conditions in the experiment when the power is varied. As discussed, the variation of the resistance $R_m$ with the pressure is also expected due to the much larger changes in the matching point with the pressure. This provides confidence in the developed treatment of the power balance.

With the values of the various parameters one can also obtain an estimate for the efficiency of the power transfer to the plasma:

$$\eta = \frac{P_p}{P_{rf}} = \frac{A^2 B \sqrt{P_{rf} - P_0}}{1 + A^2 B \sqrt{P_{rf} - P_0}}. \tag{13}$$

This definition coincides with the one introduced in [19]. The power transfer efficiencies for the different pressures are shown in figure 4(b). Clearly, the efficiency increases with the RF power. This result contradicts the findings in [19] obtained in a classical ICP discharge and leads to the conclusion that the two discharge types are governed by different mechanisms. More interestingly, the efficiency also increases with decreasing pressure. This is related primarily to a reduction of the power losses in the matching unit (lower values of $R_m$ in table 1). It could even be speculated that the better power coupling efficiency at lower pressures is a direct consequence of the increased performance of the stochastic heating. The behaviour of the plasma parameters, discussed further on, is also in support of this conclusion.

### 3.2. Spatial distribution of the induced electric field

The penetration depth of the induced electric field, i.e. the skin depth $s$, can be obtained experimentally from the modulation of the emission intensity (see section 2.3). This is based on the fact that the modulation $\chi(t)$ in the intensity is caused by the (weak) oscillation of the energy of the electrons within the RF cycle, which in turn is determined by the value of the induced electric field, $E_{\text{ind}}$.

The emission intensity of the neon line originating from the $2p_1$ state has been measured temporally resolved. A two-dimensional spatial map of the electric field is inferred, following the rationale given in section 2.3. Figure 5 shows the amplitude and the phase of the electric field of the antenna array. The first notable feature is the fact that the electric field and the phase are essentially homogeneous in a plane parallel to the array, i.e. in the vertical direction in (figure 5). However, it has to be mentioned that the camera has a finite observation angle. Consequently, due to the related collection cone the spatial resolution is changing with distance to the camera. The length scale in the figures is then defined to correspond to the one in the middle of the discharge (halfway between the front and back walls). In conclusion, the resolution is probably not better than a centimetre.

The evolution of the electric field and its phase along the horizontal direction is shown in the insets below the figures. The amplitude of the electric field (figure 5(a)) decays nearly exponentially with a skin depth ($1/e$-distance) at this conditions of about 3.8 cm. This value does not change significantly with the discharge conditions (pressure and power) in agreement with the predictions of theory [10]. A pressure-independent skin depth is also in conformity with a pressure independent value for the fitting parameter $B$, as discussed later in section 3.5. In brief, for a diffusion profile the edge density is effectively independent of the pressure, although the centre density varies linearly.

The phase of the electric field (figure 5(b)) is also changing with the distance from the antenna array. This change in the phase is related to the deposition of energy into the plasma. Consequently, the induced electric field can be treated as an evanescent wave with a finite phase velocity and the energy of the wave being absorbed by the plasma. Under the conditions of such low pressures (neon discharge at 3 Pa for the data in figure 5) this energy absorption is due mostly to the stochastic mechanism and to a lesser extent to the normal collisional absorption [10].

### 3.3. Operation in H-mode

An essential condition for the parallel stochastic heating process is a large mean free path length of the electrons $\lambda$ (longer than about the cell size $\Lambda$), i.e. collisionless electrons. This requires that for argon the neutral gas pressure $p \ll 3$ Pa. A lower gas pressure in the common inductively coupled discharges is generally related to a lower plasma density which makes the transition to the H-mode more challenging. Therefore, it is of great importance to verify the mode of operation and the mode transition power, which should be strongly dependent on $p$ [20–22].

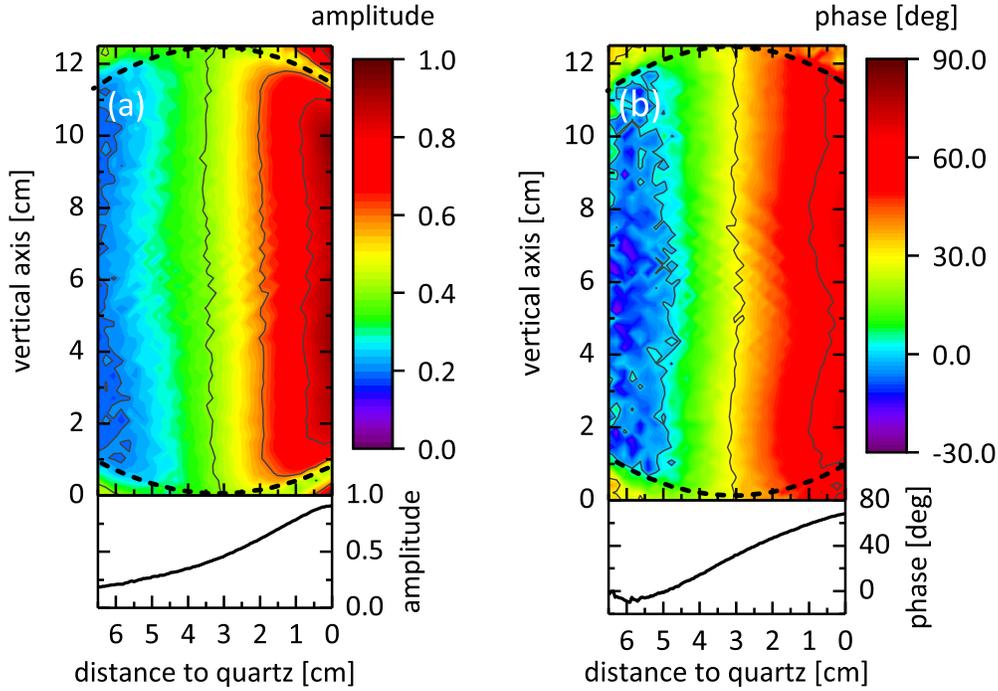

**Figure 5.** Two-dimensional spatial distribution of the amplitude in (a) and the phase in (b) of the induced electric field in neon at a pressure of 3 Pa and applied power of 400 W. Langmuir probe measurements provide $n_e = 1.3 \cdot 10^{16}\,\mathrm{m}^{-3}$ and $T_e = 4.8\,\mathrm{eV}$ at these conditions. Black dotted curves indicate the edges of the window. The insets below the contour plots show a cross section between 2 and 10 cm. The antenna array is to the right of the figures with the quartz plate from the sandwich structure at 0 cm. The zero for the vertical axis is chosen arbitrarily.

In figure 6(a) the temporally averaged plasma emission in argon is plotted as a function of the applied power from the RF generator for five different pressures between 0.5 Pa and 20 Pa. The power level for the mode transition is identified by the sudden increase in the slope of the plasma emission. Evidently, the power at which the transition occurs increases with decreasing pressure, as expected. However, even at 0.5 Pa the transition power is at approximately 150 W. This value is relatively low (power density $\sim 0.17\,\mathrm{W/cm^2}$) as compared to standard ICP discharges [23, 24] and only a factor of 3 lower than at 20 Pa.

The transition from E- to H-mode becomes even more evident, when the modulation of the emission intensity is considered (figure 6(b)). When the power level is low (40 W for the case in figure 6(b)) the discharge operates in the capacitive mode. In this mode the electrons are heated by the sheath expansion in front of the antenna. This leads to a characteristic single maximum in the plasma emission, which is well known from previous investigations of standard inductive and capacitive discharges [14, 25]. At higher power levels, above the threshold for the mode transition, the inductive coupling dominates the heating of the electrons. Under such conditions, the electrons gain energy twice per RF period. Consequently, the electron energy and the emission intensity show modulation at twice the RF frequency. This is clearly visible in the nearly perfect sinusoidal waveform at a frequency of $2\omega_{\mathrm{rf}}$, again in agreement with observations in other types of inductive discharges [14, 25].

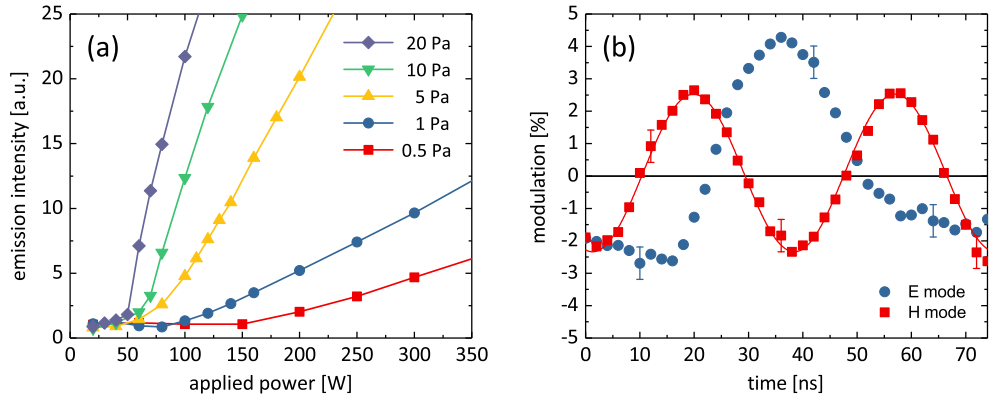

**Figure 6.** (a) Variation of the plasma emission with the power at different pressures in argon. The curves are only a guide to the eye. The uncertainties are of the order of the symbol size. (b) Temporal modulation of the emission intensity of the neon 585.25 nm line in a neon discharge at 3 Pa for two power levels: 40 W (E mode) and 400 W (H mode). The curve is a sinus fitted to the data points at 400 W.

Please note, that this oscillatory behaviour is not in a contradiction with a stochastic heating mechanism since not all electrons participate equally in the stochastic heating. Here only very energetic electrons can be observed (threshold energy for excitation 18.97 eV [16]). In fact the oscillatory behaviour is not related to a heating at all since the net energy gain over one RF period is actually zero for those electrons.

### 3.4. Spatial characteristics of the plasma

The spatial distribution of the plasma parameters is related to a variety of aspects concerning the physics of the discharge mechanism and the homogeneity in large area processing. The distribution of the EEPF and of the plasma parameters have been investigated by Langmuir probe measurements in discharges in argon and neon. Here, only the results in argon will be presented, since the trends in both gases are the same. Measurements are carried out between the centre of the discharge and the wall (range of 21 cm). The results are complemented by spatially resolved optical emission measurements along the diagonal of the antenna array.

From the probe measurements the plasma potential and the energy distribution function of the electrons are obtained from the second derivative, i.e. by the Druyvesteyn method. The electron density and (effective) temperature are then calculated from the EEDF through integration, as discussed in section 2.3. For easier comparison with a Maxwellian distribution, it is common to present the EEPF instead of the EEDF. The obtained EEPF at all positions are close to Maxwellian distributions (figure 7). The reason for this Maxwellian character is not yet fully understood and is a topic of an ongoing investigation. Clearly, Coulomb collisions at the densities of $10^{16}$ m$^{-3}$ are too rare to allow for thermalization. It can be speculated that the Maxwellian character is related to the stochastic heating mechanism, as discussed in more detail in the next section.

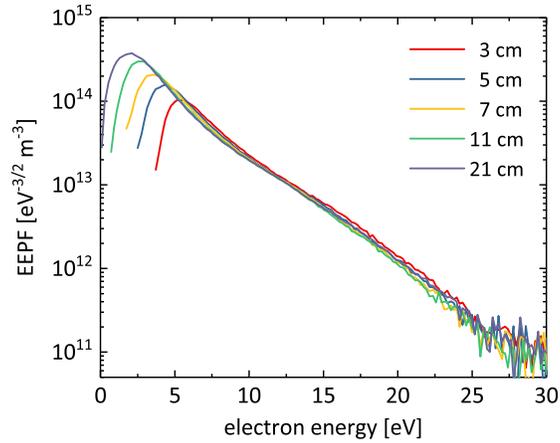

**Figure 7.** EEPF obtained at different probe positions in Ar at 0.5 Pa and 300 W. Each curve has been shifted by the local value of the plasma potential.

A further characteristic of the obtained distributions is their non-locality. This is a direct consequence of the low working pressure, required for the stochastic heating mechanism. With diminishing pressure, the electron mean free path and the energy relaxation length, $\lambda_\varepsilon$, become comparable to or even larger than the dimensions of the plasma. To demonstrate this, the mean free path (mfp) of the electrons for elastic and inelastic collisions, $\lambda$ and $\lambda^*$, respectively, together with the energy relaxation length have been calculated for an argon discharge at 0.5 Pa. The density of the argon atoms has been obtained from the pressure and the measured value of the gas temperature ($T_g = 350$ K) and the collision cross sections are taken from [26]. The results are presented in figure 8. The mfp for elastic collisions is of the order of a few cm, whereas the mfp for inelastic collisions is larger than 40 cm. The energy relaxation length in the elastic energy range ($\varepsilon < 11.55$ eV) is several meters, well beyond the size of the plasma chamber, while in the inelastic energy range ($\varepsilon > 11.55$ eV) it is between 1 m and 20 cm, i.e. comparable with the chamber dimensions. Under such conditions the EEDF no longer depends on the position and the kinetic energy separately, but is a unique function of the total energy of the electrons, kinetic and potential. This is best visualized by shifting the energy axis of the measured EEDF by the local value of the plasma potential [15, 27, 28]. When the distribution function is non-local, the shifted distributions, measured at different positions, fall on the same unique curve. As expected, this is well fulfilled under the experimental conditions here (figure 7). Obviously, non-locality comes along with the collisionless requirement for the stochastic heating mechanism in INCA.

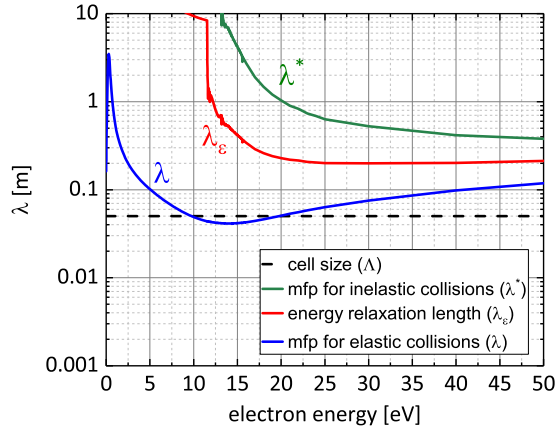

**Figure 8.** Mean free path (mfp) for elastic, $\lambda$, and for inelastic collisions, $\lambda^*$, in argon at 0.5 Pa and $T_g = 350\,\mathrm{K}$. The energy relaxation length $\lambda_\varepsilon$ for the electrons and the cell size $\Lambda$ of the array are also shown.

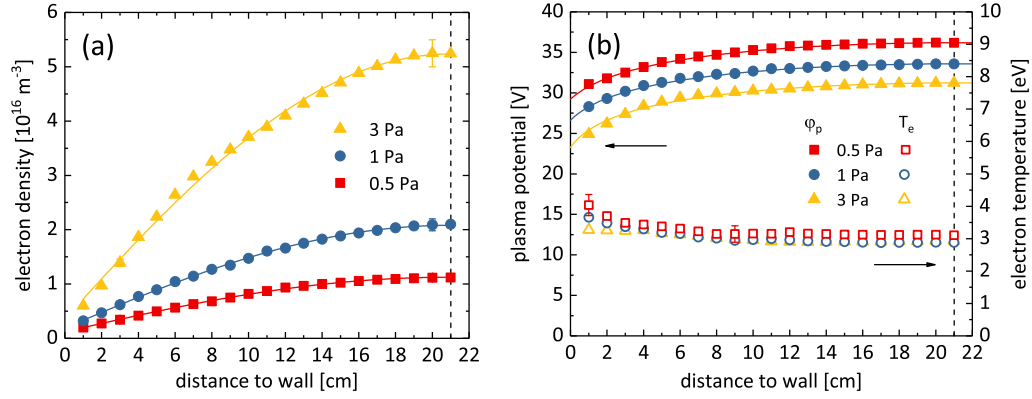

**Figure 9.** Spatial profiles of the electron density in (a) and of the plasma potential and the electron temperature in (b) measured in Ar at 300 W and various pressures. The uncertainties are of the same order as the size of the symbols.

From the distribution functions, the plasma density $n_\mathrm{e}$ and the electron temperature $T_\mathrm{e}$ are obtained (figure 9). The electron density (figure 9(a)) has an ambipolar diffusion profile for all investigated conditions. This has been emphasized in the figure by adding on top of the experimental data the theoretical profile in one dimension:

$$n(x) = n_0 \cos\left[(1-\delta)\frac{\pi}{L}\left(x - \frac{L}{2}\right)\right]. \tag{14}$$

Here, $n_0$ is the plasma density in the center, $L$ is the chamber width and $\delta$ is a fitting parameter, that accounts for the boundary conditions at the sheath edge. In [29] it has been shown theoretically that this parameter is related to the ion collisionality through the ion-neutral collision frequency $\nu_\mathrm{mi}$:

$$\delta = \frac{u_\mathrm{B}}{L\nu_\mathrm{mi}} \ll 1. \tag{15}$$

Further, it defines the edge-to-centre ratio of the plasma density:

$$\frac{n_{\mathrm{s}}}{n_0} = \cos\left((1-\delta)\frac{\pi}{2}\right) \approx \frac{\pi}{2}\delta. \tag{16}$$

Note then, that this ratio is inversely proportional to the ion-neutral collision frequency, i.e. to the neutral gas pressure.

A prerequisite for the appearance of such a profile is a homogeneous ionization profile [1]. The excellent agreement between the theoretical profile and the measurement reveals that the ionization profile in INCA is indeed homogeneous. This aspect is further confirmed by the profile of the electron temperature (figure 9(b)). A flat, nearly constant profile of $T_{\mathrm{e}}$ is observed. Both the homogeneity of the ionization profile and of the electron temperature are related to the non-locality and the nearly Maxwellian nature of the EEDF.

Figure 10 shows the emission intensity over the diagonal of the discharge. The intensity has been integrated in the wavelength range between 400 and 900 nm. However, as the inset shows, the strongest lines lie in the 700 to 900 nm region. These are lines originating from the 2p manifold of the argon atom and are typically the most intensive ones. The data are rather noisy, since they had to be corrected for reflection from the copper coils of the array. The emission profiles (figure 10) resemble more or less the diffusion profiles of the plasma densities (figure 9(a)). This is not surprising since the distribution function is close to a Maxwellian and the emission intensity for ground state excitation is proportional to the plasma density. The difference between the two profiles is mostly due to the difference in the value of $\delta$. As discussed, this parameter scales inversely proportional to the gas pressure. Consequently, at higher pressures the density near the walls relatively to the centre density is much lower. For the corner points along the diagonal the scaling is even stronger – the edge-to-centre density ratio depends on $\delta^2 \propto p^{-2}$.

The profile of the plasma potential has been also obtained (figure 9(b)). Due to the homogeneous profile of the electron temperature, it is directly related to the profile of the electron density through the Boltzmann relation:

$$e\phi = e\phi_0 + k_{\mathrm{B}}T_e \ln\left(\frac{n}{n_0}\right). \tag{17}$$

Here, $e$ is the elementary charge and $k_{\mathrm{B}}$ is the Boltzmann constant. Using the same parameters from the profiles of the electron density and the measured values of the electron temperature, the expected profiles of the plasma potential are also presented in figure 9(b). The only adjustable parameter here is the value of the potential in the centre, $\phi_0$. The good agreement between measurements and theoretical expectations shows the consistency of the profiles of the different quantities.

However, it has to be noted that the values of the potential in the centre $e\phi_0$ and at the sheath edge $e\phi_m s$ are much higher than the value for the floating potential expected from the well-known relation

$$e\phi_{\mathrm{fl}} = \frac{k_{\mathrm{B}}T_e}{2} \ln\left(\frac{m_{\mathrm{i}}}{2\pi m_{\mathrm{e}}}\right). \tag{18}$$

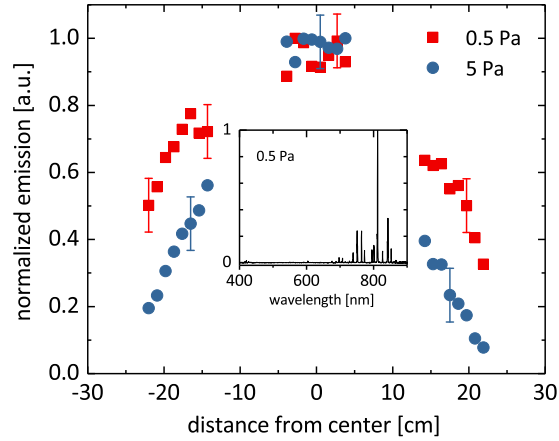

**Figure 10.** Spatial profiles (along the diagonal of the antenna-facing wall) of the emission intensity (integrated over the range 400-900 nm) in Ar at 300 W. The grouping of the points is due to the availability of viewing ports. The errorbars represent the reproducibility of the data. The inset shows the spectrum at the centre for the 0.5 Pa case.

Here, $m_i$ is the ion mass and $m_e$ the mass of an electron. For an argon plasma with an electron temperature of 4 eV the potential close to the walls should be about $\phi_{fl} = 19$ V. This value is much lower than the measured near-wall potential of $\phi_s = 30$ V.

The obtained high values ($\phi_0 = 36$ V at 0.5 Pa) have been consistently confirmed by the energies of the ions hitting the wall that were measured independently by a retarding field energy analyzer (see also next section). Such values of the plasma potential are possible only if a group of super energetic electrons are present in the discharge. When such electrons are present, Demidov *et al* [30] have demonstrated, that the value of the plasma potential is increased drastically, even if the density of these super energetic electrons $n_{ef}$ with energies higher than the confining potential is several orders of magnitudes smaller than the density of the bulk electrons $n_e$. They have shown that the value of the plasma potential is determined in this case by a modified expression:

$$e\phi_s = \frac{k_B T_e}{2} \ln\left(\frac{m_i}{2\pi m_e}\right) + k_B T_e \ln\left(\frac{j_i}{j_i - j_{ef}}\right) \tag{19}$$

that additionally accounts for the flux density $j_{ef}$ of these fast electrons. Here, $j_i$ is the flux density of the ions to the wall. When $j_{ef} \approx j_i$ then $\phi_s \gg \phi_{fl}$ and the sheath potential is no longer related to the electron temperature. Instead, the energy of the fast electrons determines the value of the potential. When these electrons have only a small energy spread, i.e. when they are nearly mono-energetic, then the value of the sheath potential is a good measure of their energy [30].

Such super energetic electrons are expected to be present in the discharge due to the resonance nature of the heating mechanism which forms the basis for the operation of the INCA discharge. Therefore, the obtained high values of the plasma potential are an indirect indication for the presence of such electrons and, hence, for the substantial role that the novel stochastic heating is playing in this discharge.

Figure 9(b) also reveals that with a decreasing pressure there is a notable increase of the electron temperature towards the edge. This could be attributed to the presence of super energetic electrons which should play a more important role close to the boundary. At the edge the density of the main electrons decreases strongly. At the same time the density of the super energetic electrons is not influenced by the confining potential and remains nearly the same. Consequently, the relative contribution of these electrons increases towards the edge.

By equation (19), an estimate for the density of the super energetic electrons can be obtained. Their flux density is given by

$$\frac{j_{\mathrm{ef}}}{j_{\mathrm{i}}} = 1 - \sqrt{\frac{m_{\mathrm{i}}}{2\pi m_{\mathrm{e}}}} \exp\left(-\frac{e\phi_{\mathrm{s}}}{k_{\mathrm{B}}T_{\mathrm{e}}}\right). \tag{20}$$

Using the measured values of the plasma potential and the electron temperature, one obtains that the second term on the right of (20) is negligible. Then the ion flux to the walls is essentially compensated by the flux of fast electrons: $j_{\mathrm{ef}} \approx j_{\mathrm{i}} = n_{\mathrm{s}}u_{\mathrm{B}}$ ($u_{\mathrm{B}}$ is the Bohm velocity). The flux of the bulk electrons is negligible and these electrons are confined in the plasma. Then, the density of the fast electrons $n_{\mathrm{ef}}$ can be estimated as:

$$n_{\mathrm{ef}} \approx n_{\mathrm{s}}\frac{u_{\mathrm{B}}}{u_{\mathrm{ef}}} \approx 2.7 \cdot 10^{12}\,\mathrm{m}^{-3}. \tag{21}$$

Here the velocity of the fast electrons $u_{\mathrm{ef}}$ is estimated by assuming that these electrons are nearly mono-energetic with an energy given by the value of the sheath potential $e\phi_{\mathrm{s}} = 30\,\mathrm{eV}$. Accidentally, this value coincides with the energy at which the EEDF obtained by the Langmuir probe enters the noise region. A value for the plasma density near the sheath of $n_{\mathrm{s}} = 2 \cdot 10^{15}\,\mathrm{m}^{-3}$ and an electron temperature (near the sheath) of $k_{\mathrm{B}}T_{\mathrm{e}} = 4\,\mathrm{eV}$ are used. These values are taken from figure 9 for the case at 0.5 Pa and 300 W. Apparently the density of these fast electrons is about three orders of magnitude lower than the bulk density making their direct observation in the Langmuir probe measurements a challenge. However, the values of the plasma potential are a sensitive indicator for their presence.

### 3.5. Influence of power and pressure

The presence of a group of super energetic electrons and the efficiency of the novel stochastic heating mechanism become even more evident in the variations of the plasma parameters with the applied power and the neutral gas pressure. Here, all measurements with the Langmuir probe are performed in the centre of the discharge and the measurements of the IVDF are performed at the centre of the antenna-facing wall of the discharge.

For all pressures, the electron density increases linearly with the power from the RF generator (figure 11(a)), with almost constant zero crossing at $P_{\mathrm{rf}} = P_0 \approx 100\,\mathrm{W}$. This value corresponds well to the data from the optical emission measurements at which a transition into the H-mode is observed (see figure 6(a)). Although theoretically a linear scaling of the density with the power is expected [1], this should be the power deposited into the plasma and not the power from the generator. Using the obtained power deposition efficiencies from figure 4, the densities in figure 11(a) can be plotted as a function of $P_p$ as well. The result is shown in figure 11(b). The expected proportionality is well reproduced and in accordance with the general theoretical expectations and measurement results in standard ICP discharges [31–34].

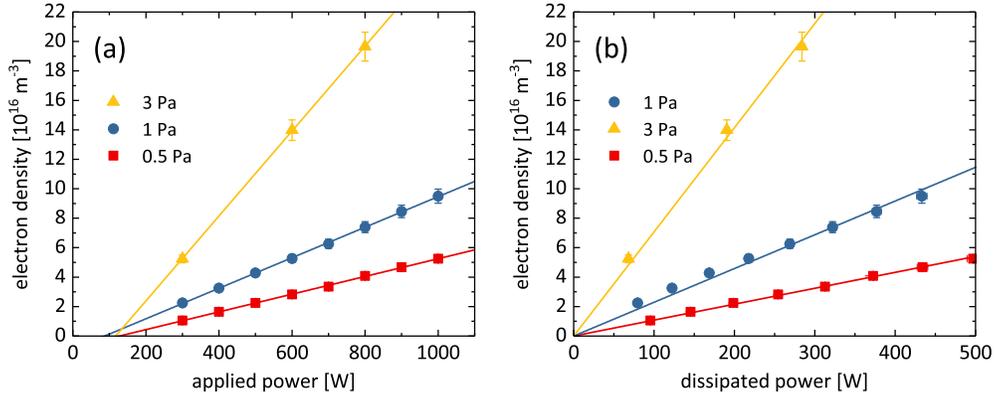

**Figure 11.** (a) Variation of the electron density in the centre of the plasma chamber with the applied power at several values of the pressure. The lines are linear fits to the data. (b) The same data as a function of the power dissipated in the plasma $P_p$. The straight lines give the expected linear dependence.

This behaviour can be quantified in the following way [1]. From the power balance in a stationary plasma it follows that the deposited power is equilibrated by the power lost, $P_l$: $P_p = P_l$. The latter can be expressed as [1]:

$$P_l = \varepsilon_t \nu_{iz} V \bar{n}_e. \tag{22}$$

Here, $\nu_{iz}$ is the ionization frequency, i.e. the rate at which electrons are produced, and $V$ is the plasma volume. The volume-averaged density, $\bar{n}_e$, is related to the density in the centre $n_0$ by (diffusion profile):

$$\bar{n}_e \approx \left(\frac{2}{\pi}\right)^3 n_0. \tag{23}$$

On the other hand, the energy required for the creation of an electron-ion pair, $\varepsilon_t$, includes the energy loss due to collisions, $\varepsilon_c$, taken from [1], the energy for sustaining the electrostatic potential, $e\phi_s$, as well as the kinetic energy of the electrons. This kinetic energy is comprised only of the energy in the movement in the two directions parallel to the walls, $2(k_B T_e/2)$. Then $\varepsilon_t = \varepsilon_c + k_B T_e + e\phi_s$.

Note, that using the continuity equation, the product $\nu_{iz} V \bar{n}_e$ in (22) can also be expressed through the Bohm flux averaged over the plasma surface area $A$. However, in this case the value of the density at the sheath edge, $n_s$, needs to be known. Since measurements near the sheath edge are challenging, this value can only be estimated from the present measurements. Therefore, using the Bohm flux provides a less accurate estimation than the one used in (22).

Using (22) and the measured electron and gas temperatures, the slope

$$\frac{\partial n_0}{\partial P_{\mathrm{p}}} = \left(\frac{\pi}{2}\right)^3 \frac{1}{\varepsilon_{\mathrm{t}} \nu_{\mathrm{iz}} V} \tag{24}$$

can be estimated. However, $\nu_{\mathrm{iz}}$ is rather sensitive to even slight inaccuracies in the electron temperature. Furthermore, the exact shape of the EEDF in the inelastic energy range is also strongly influencing the values. Therefore, it is more accurate to express the ionization frequency from the continuity equation using the diffusion profile of the density:

$$\nu_{\mathrm{iz}} = \frac{D_{\mathrm{a}}}{\Lambda_{\mathrm{d}}^2}. \tag{25}$$

Here $D_{\mathrm{a}} \approx k_{\mathrm{B}} T_{\mathrm{e}} / (m_{\mathrm{i}} \nu_{\mathrm{mi}})$ is the ambipolar diffusion coefficient. The classical diffusion length $\Lambda_{\mathrm{d}} = 3.79\,\mathrm{cm}$ depends only on the geometry and the dimensions of the plasma volume. Then the slope of the plasma density (24) becomes:

$$\frac{\partial n_0}{\partial P_{\mathrm{p}}} = \left(\frac{\pi}{2}\right)^3 \frac{\Lambda_{\mathrm{d}}^2}{V} \frac{m_{\mathrm{i}} \nu_{\mathrm{mi}}}{k_{\mathrm{B}} T_{\mathrm{e}} \varepsilon_{\mathrm{t}}}. \tag{26}$$

The first two ratios are constants that depend only on the geometry and the dimensions of the plasma volume. Then it becomes evident, that when the electron temperature is nearly constant, the slope of the density is independent of the absorbed power and scales linearly with pressure.

Using relation (26), a comparison between the experimental slopes in figure 11(b) and the theoretical expectations can be made. The values are provided in table 2. Very good agreement between the two values is obtained with a deviation of about 20 %. Further, one can also use the measured values of the electron and gas temperatures to verify the validity of (25). The right-hand side is readily obtained with the experimental values in table 2. The left-hand side – the effective ionization frequency for a non-homogeneous distribution function – is given by:

$$\nu_{\mathrm{iz}}^{(\mathrm{eff})} = \frac{\int \nu_{\mathrm{iz}} n_{\mathrm{e}} \mathrm{d}V}{\int n_{\mathrm{e}} \mathrm{d}V}. \tag{27}$$

Here, this average is not explicitly carried out but for simplicity the limiting values $\nu_{\mathrm{iz}}^{(\mathrm{max})} > \nu_{\mathrm{iz}}^{(\mathrm{eff})} > \nu_{\mathrm{iz}}^{(\mathrm{min})}$ are estimated from the values of the electron temperature and the ionization rate coefficient $K_{\mathrm{iz}}(T_e)$ shown in [1, p. 80]. The values determined for the diffusion rate fall within these limits except for the highest pressure at $p = 3\,\mathrm{Pa}$. However, there the distribution function is no longer Maxwellian and the depleted tail (see figure 13 further down) indeed leads to a reduced value. Still, this shows that the various quantities are consistent with each other and that the discharge behaviour is well described by the presented models.

The power absorbed by the plasma, $P_{\mathrm{p}}$, can also be expressed through the energy that the electrons gain from the array. In this case $P_{\mathrm{p}}$ is determined by the energy that an electron gains on average in the skin layer, $\langle\Delta\varepsilon\rangle$: $P_p = \langle\Delta\varepsilon\rangle\nu_h$. The rate at which electrons are entering this layer is given by

$$\nu_h = \frac{v_{\mathrm{th}}}{4}\int n(x,y,s)\,\mathrm{d}A \approx n_0 v_{\mathrm{th}}\left(\frac{L}{\pi}\right)^2 \sin^2\left[\frac{\pi}{2}\left(1-\frac{2a}{L}\right)\right]\cos\left[\frac{\pi}{2}\left(1-\frac{2s}{D}\right)\right]. \quad (28)$$

**Table 2.** Comparison of the slopes of the plasma density in the centre with the absorbed power $P_p$ obtained from the experiment (figure 11(b)) and from the theory (equation (26)). The values of the relevant quantities for the calculation are also provided. The values of $\varepsilon_c$ are taken from [1, p. 82] and for the ambipolar diffusion coefficient the mobility of argon ions in their parent gas at STP $b_0 = 1.54\,\mathrm{cm^2/(V\,s)}$ from [35] is used. For the gas temperature the measured value of $T_g = 350\,\mathrm{K}$ has been used.

| $p$ [Pa] | | 0.5 | 1 | 3 |
|---|---|---|---|---|
| $\partial n_0/\partial P_p|_{\mathrm{exp}}$ | [$10^{14}\,\mathrm{W^{-1}\,m^{-3}}$] | 1.1 | 2.3 | 7.1 |
| $\partial n_0/\partial P_p|_{\mathrm{th}}$ | [$10^{14}\,\mathrm{W^{-1}\,m^{-3}}$] | 1.3 | 2.9 | 9.9 |
| $\nu_{\mathrm{iz}} = D_a/\Lambda_d^2$ | [$10^4\,\mathrm{s^{-1}}$] | 10.3 | 4.50 | 1.33 |
| $\nu_{\mathrm{iz}}^{(\mathrm{min})} = N_g K_{\mathrm{iz}}(T_e^{(\mathrm{min})})$ | [$10^4\,\mathrm{s^{-1}}$] | 3.0 | 3.2 | 9.3 |
| $\nu_{\mathrm{iz}}^{(\mathrm{max})} = N_g K_{\mathrm{iz}}(T_e^{(\mathrm{max})})$ | [$10^4\,\mathrm{s^{-1}}$] | 10.4 | 11.5 | 18.6 |
| $T_e^{(\mathrm{min})}$ | [eV] | 3.1 | 2.9 | 2.9 |
| $T_e^{(\mathrm{max})}$ | [eV] | 4 | 3.5 | 3.1 |
| $\phi_s$ | [V] | 30 | 27.5 | 24 |
| $\varepsilon_c$ | [eV] | 43 | 50 | 53 |
| $\varepsilon_t$ | [eV] | 78 | 81 | 80.1 |
| $D_a$ | [$\mathrm{m^2/s}$] | 147.6 | 64.6 | 19.1 |

Here, the diffusion profile (14) has been used (neglecting the boundary contribution $\delta \ll 1$). Further, it has been assumed that the heating takes place only in the volume above the antenna. In our case, the edge coils of the antenna are a distance of $a = 5\,\mathrm{cm}$ away from the walls. The depth of the plasma volume is $D = 130\,\mathrm{mm}$. Then, for a skin width of $s = 3.5\,\mathrm{cm}$, the rate is $\nu_h/(n_0 v_{\mathrm{th}}) = 1.16 \cdot 10^{-2}\,\mathrm{m^2}$. Using the values at $P_{\mathrm{rf}} = 800\,\mathrm{W}$ (figure 11), one obtains from the experiment $\langle\Delta\varepsilon\rangle = 3.7\,\mathrm{eV}$, $1.8\,\mathrm{eV}$ and $0.72\,\mathrm{eV}$ at $0.5\,\mathrm{Pa}$, $1\,\mathrm{Pa}$ and $3\,\mathrm{Pa}$, respectively. It is evident that the energy gained decreases with the pressure. Equation (26) predicts that $\langle\Delta\varepsilon\rangle$ scales inversely proportional to pressure for a constant electron temperature. The reason is the more efficient plasma production at higher pressure due to the decrease in particle losses. Then, the same energy has to be distributed among a larger number of electrons, leading to a smaller energy gain per electron.

As was already discussed, the slope of the plasma density with power $\partial n_0/\partial P_p$ scales linearly with pressure (equation (26)). This means that at a constant power the density in the plasma centre $n_0$ should scale linearly with pressure. This behaviour is indeed observed in the experiment, both for argon and neon (figure 12(a)). The plasma density increases linearly with the pressure and has a zero crossing at 0 Pa, i.e. $n_0 \propto p$. Similar results have been observed also in earlier works on conventional ICP discharges [36,37]. In our case the result may seem to contradict on first sight the observed independence of the skin depth $s$ on the pressure (as obtained from the analysis of the modulation of the emission intensity). The variation of the skin layer width is contained also in the parameter $B$ in section 3.1. Therefore, the constancy of $s$ with the pressure is further confirmed by the constant (pressure-independent) value of $B$.

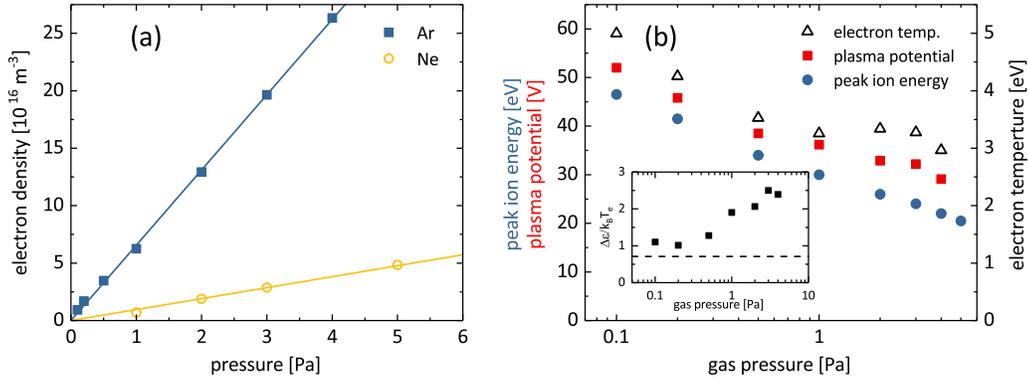

**Figure 12.** (a) Variation of the electron density in the center of the plasma chamber with pressure in discharges in Ne and Ar at $P_{rf} = 800\,\text{W}$. The lines are linear fits through the points. (b) Variation of the electron temperature and the plasma potential with pressure in Ar at $P_{rf} = 800\,\text{W}$. The energy of the peak in the IVDF is also shown as a measure for the potential drop over the sheath. The uncertainties in the potential and in the energy of the peak are of the order of the symbol size. The inset shows the potential drop over the plasma bulk normalized to the electron temperature. The dashed line shows the theoretical limit of $1/\sqrt{2}$.

The apparent contradiction is resolved if one considers that the skin depth depends on the plasma density in front of the antenna, i.e. the density at the sheath edge $n_s$, whereas equation (26) and figure 12(a) refer to the density in the centre of the discharge. Since the edge-to-centre plasma density ratio (see equation (16)) scales inversely proportional with pressure [1, 29], this compensates exactly the increase in the centre density with the gas pressure. Consequently, the plasma density near the sheath edge remains constant, leading to the observed constant field penetration depth and to the obtained constant value of the parameter $B$.

The pressure dependence of the plasma potential and of the electron temperature show a decreasing trend (figure 12(b)). Note that in the pressure range 0.5 to 3 Pa the temperature remains nearly constant (figure 9(b)). The plasma potential in the centre, $\phi_0$, mirrors the temperature behaviour, but the peak ion energy, dominantly determined by the sheath potential $\phi_s$, lacks the saddle point (figure 12(b)).

To understand the different behaviour of the two potentials, $\phi_0$ and $\phi_s$, one has to take a look at the potential drop over the plasma bulk, given by their difference $\Delta\varepsilon = e(\phi_0 - \phi_s)$. The results show a consistent increase of this potential drop with the pressure (inset of figure 12(b)). This is a well known and expected behaviour [14]. It is also connected through the Boltzmann relation to the edge to centre plasma density ratio. The increase in the pressure leads to higher ion friction through collisions with the background gas. In order to reach the Bohm velocity at the sheath edge, the ions now need a higher ambipolar field/more energy, which is supplied by increasing the potential drop over the plasma bulk. A higher ambipolar field is directly projected into higher density gradients over the plasma bulk and into lower edge to centre plasma densities. Theoretically, in the range of low pressures when the ion friction becomes negligible, the potential drop over the plasma bulk has to saturate to $k_\mathrm{B}T_\mathrm{e}/\sqrt{2}$. Indeed, such saturation is observed below 0.5 Pa when the ion mean free path becomes comparable with the shortest plasma dimension $D$. However, the saturation value is not $k_\mathrm{B}T_\mathrm{e}/\sqrt{2}$, but $k_\mathrm{B}T_\mathrm{e}$. This is probably a consequence of the way how this potential drop was estimated. As shown in [15] the peak energy of the IVDF does not exactly coincide with the sheath potential.

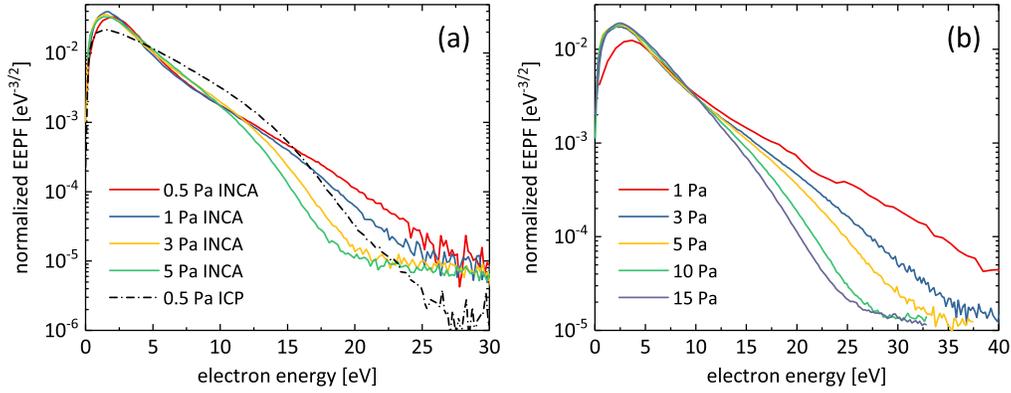

**Figure 13.** Variation of the EEPF in the center of the plasma chamber with pressure in (a) argon plasma at 300 W and (b) neon plasma at 800 W. The dash-dotted line in (a) shows the EEPF measured in a conventional ICP [38].

Probably the most convincing evidence for the role of the stochastic heating in INCA is the variation of the electron distributions with pressure. Figure 13 presents these variations measured in the centre of the discharge. For a better comparison, the EEPFs have been normalized to the electron density, i.e. to the corresponding integral over the EEPF itself (see equation (3)). In this representation, a Maxwellian distribution is given by:

$$\frac{1}{n_\mathrm{e}}f_\mathrm{p}(\varepsilon) = \frac{2}{\sqrt{\pi(k_\mathrm{B}T_\mathrm{e})^3}} \exp\left(-\frac{\varepsilon}{k_\mathrm{B}T_\mathrm{e}}\right). \tag{29}$$

Indeed, as discussed above the electron density of the fast electrons is too low to be detected by the Langmuir probe. However, the interaction of the low-energy electrons with the heating field is still visible in the EEPF shape. More specifically, it is the electrons in the inelastic energy range that show sensitive dependence on the pressure. These electrons have sufficient energy to excite and ionize the atoms of the background gas. In inelastic collisions they lose most of their energy. Naturally, this sink of electron energy is always present. However, the population of these electrons is very sensitive to the presence and the efficiency of an energy source, since they have the shortest energy relaxation length $\lambda_\varepsilon$ (see figure 8).

The sensitivity of the electron population in the inelastic energy range is very well visible in the results in figure 13. At low pressures ($\leq 1\,\mathrm{Pa}$ for argon and $\leq 5\,\mathrm{Pa}$ for neon) the EEPF is nearly Maxwellian with a clear presence of electrons in the inelastic energy range ($\varepsilon > 11.55\,\mathrm{eV}$ for argon and $\varepsilon > 16.6\,\mathrm{eV}$ for neon). These limiting values of the pressure correspond to an electron energy relaxation length in the inelastic range of about $0.1\,\mathrm{m}$ for both gases (for argon, see figure 8). This is about the length $D$ of the short plasma dimension. Above these pressures values, the low energy electrons retain their Maxwellian character since the energy relaxation lengths in the elastic energy range are still much longer than the chamber dimensions. However, a noticeable depletion of electrons in the inelastic energy range is observed (figure 13). Such a depletion is typically observed also in conventional ICP [38]. However, there it is seen even at low pressures (figure 13(a)). The conclusion to be made here is that in INCA a heating mechanism operates, that efficiently heats the electrons and restores their population in the inelastic energy range. At higher pressures, the collisions diminish its efficiency and simultaneously make the distribution of electrons in the inelastic energy range local. Consequently, the EEPF returns to its more conventional shape, typical for ICPs.

## 4. Conclusion

In this study, first experimental results on the investigation of the <u>IN</u>ductively <u>C</u>oupled <u>A</u>rray (INCA) discharge are presented. In INCA the plasma is created by an array of 36 small inductive coils. This design realizes experimentally for the first time a theoretical concept for stochastic electron heating at low pressures, proposed recently [9, 10]. Efficient plasma operation in INCA at pressures down to $0.1\,\mathrm{Pa}$ is demonstrated as well as an almost homogeneous electron excitation over the whole antenna array surface. The power levels for the transition into the H-mode do not exceed $150\,\mathrm{W}$. The corresponding power density for the transition (about $0.17\,\mathrm{W/cm^2}$) is unusually low for a conventional inductive discharge. The plasma parameters, e.g. the unusually high plasma potential and the energy distribution functions of the electrons and the ions consistently indicate the presence of super energetic electrons (energies of about several tens of eV). These results provide strong experimental indications for the existence and the efficiency of the proposed novel stochastic heating mechanism.

The results are initial experimental investigations of this type of source and certainly different aspects of the INCA discharge can be further improved. Despite the extensive experimental results obtained, a number of open questions and challenges still remain. The reason for the Maxwellian nature of the EEDF under collisionless conditions and the direct detection of the resonant super energetic electrons are probably the most important outstanding issues. Other important aspects, especially for processing applications, concern the operation in molecular gases with increased collisional losses and presence of negative ions. Currently, the experimental efforts concentrate on imposing a magnetic cusp field to reduce the electron losses to the walls as well as on the spatial structure of the induced electric field of the antenna array.

However, already at this stage of the investigations it is clear that this discharge concept offers interesting opportunities both for theoretical studies and for practical applications. Prominent application examples include the use of INCA for large area plasma processing or its utilization as a thruster for space propulsion.

## Acknowledgments

Financial support from the Research School at Ruhr-University Bochum is gratefully acknowledged. The authors are indebted to Th. Zierow, F. Kremer, B. Becker and the mechanical workshop at the Faculty of Physics and Astronomy, Ruhr-University Bochum for the expert technical assistance.